\documentclass[12pt,sort&compress]{elsarticle}

\usepackage{graphicx}
\usepackage[left=2.5cm,right=2.5cm,top=2.5cm,bottom=2.5cm]{geometry}

\usepackage[section]{placeins}

\usepackage{enumitem}
\setlist{noitemsep}

\usepackage{xcolor}
\usepackage{soulutf8}
\definecolor{lightpink}{rgb}{1,.90,.90}
\definecolor{lightred}{rgb}{1,.50,.50}
\sethlcolor{yellow}

\definecolor{commentsColor}{rgb}{0.497495, 0.497587, 0.497464}
\definecolor{keywordsColor}{rgb}{0.000000, 0.000000, 0.635294}
\definecolor{stringColor}{rgb}{0.558215, 0.000000, 0.135316}

\usepackage{tcolorbox}
\tcbuselibrary{breakable,listings,skins}

\newtcblisting{sourcecodewide}[1]{enhanced jigsaw, left skip=-5mm, right skip=-2mm,
             colback=yellow!10,
             colbacktitle=blue!15!white,
             colframe=yellow!50!black,
             coltitle=black,
             boxsep=0mm,
             breakable, listing only, title=#1, fonttitle=\bfseries,  
             listing engine=listings, listing options={language=c++, gobble=0, firstline=0, numbers=left, framesep=2mm, keepspaces=true, numberstyle=\ttfamily\tiny, basicstyle=\ttfamily, extendedchars=true, keywordstyle=\ttfamily\color{keywordsColor}\bfseries, commentstyle=\color{commentsColor}\ttfamily\textit, stringstyle=\ttfamily\color{stringColor}, columns=flexible}}

\newtcblisting{sourcecode}[1]{enhanced jigsaw,
             colback=yellow!10,
             colbacktitle=blue!15!white,
             colframe=yellow!50!black,
             coltitle=black,
             boxsep=0mm,
             top=0mm,
             bottom=0mm,
             left=15pt,
             breakable, listing only,
             title=#1, fonttitle=\bfseries,  
             listing engine=listings, listing options={language=c++, gobble=0, firstline=0, numbers=left, framesep=2mm, keepspaces=true, numberstyle=\ttfamily\tiny, basicstyle=\ttfamily, extendedchars=true, keywordstyle=\ttfamily\color{keywordsColor}\bfseries, commentstyle=\color{commentsColor}\ttfamily\textit, stringstyle=\ttfamily\color{stringColor}, columns=flexible, framesep=0pt, rulesep=0pt, numbersep=5pt}}

\usepackage[bookmarks=true,colorlinks=true,citecolor=blue,linkcolor=blue,urlcolor=magenta]{hyperref}

\begin{document}

\begin{frontmatter}

\title{Geant4 simulation of the moderating neutrons spectrum}

\author[addr1]{V.P. Smolyar\corref{cor1}}
\cortext[cor1]{Corresponding author e-mail: svp@op.edu.ua}
\author[addr1]{V.A. Tarasov}
\author[addr1]{A.O. Mileva}
\author[addr2]{A.V. Tykhonov}
\author[addr1]{V.D. Rusov}

\address[addr1]{ Department of Theoretical and Experimental Nuclear Physics,\\
Odessa Polytechnic National University, Odessa, Ukraine}

\address[addr2]{Department of Nuclear and Particle Physics, University of Geneva, CH-1211, Switzerland}

\date{\today}


\begin{abstract}

%

%

In this paper we 
deal with the problem of predicting a steady-state neutron
spectrum in 
media 
of arbitrary composition and geometry.
The 
analytical calculations of such spectrum are often too complex, if at all
possible.
We describe a method of Geant4-based Monte Carlo calculation 
of
the steady-state neutron spectrum
in a medium containing a fixed neutron source.
In addition to the steady-state spectrum,
 we obtain the snapshots of the neutron spectrum evolution in time,
which may be thought of as the non-equilibrium neutron spectra, and 
their form is of considerable interest for further studies.
\end{abstract}


\end{frontmatter}

\section{Introduction}

%

The detailed knowledge of the neutron spectrum is crucial for numerous
applications such as the nuclear reactor operation~\cite{Ferziger1966,Williams1966,Stacey2018}, the traveling wave reactor
(TWR) development~\cite{Rusov2015Peculiarities,Feoktistov1989,Teller1996}, including the search of the neutron energy ranges suitable
for the wave nuclear burning~\cite{Rusov2011Energies,Tarasov2015Ultraslow}, the search and prediction of the so-called 
``blowup modes'' in neutron-multiplying media~\cite{Rusov2011Sharpening,Rusov2013Fukushima}, the verification of neutron
moderation theories and so on.

However, obtaining the neutron spectrum for a particular
combination of geometry, medium composition, set of nuclear reactions etc. is a
non-trivial and rather complex task by itself. Therefore, in practice, various assumptions
and approximations are usually made. For example, in nuclear
reactor physics the neutron spectrum is often
composed of
 separate parts --
the fission spectrum in high-energy range (fast neutrons), the Maxwellian
spectrum in low-energy range (thermal neutrons), and the Fermi spectrum in
between. These three parts are then joined together using some scaling
coefficients~\cite{Stacey2018}.
For the same reasons, e.g. in~\mbox{\cite{Rusov2011Energies}}, a well-known fission
spectrum was used for the fast TWR calculations. In~\mbox{\cite{Tarasov2015Ultraslow}}
the experimental WWER\footnote{Water-water energetic reactor} spectrum was used for the epithermal TWR calculations.
So there is currently no straightforward way to
obtain a complete form of the neutron spectrum within a single theory.
It is not surprising, because such theory is not at all easy to develop, and the
well known and most cited theories of neutron moderation were largely developed with nuclear reactors in mind
(``\textit{...raison d'\^{e}tre of slowing-down calculations (or experiments)
is to provide information for criticality calculations.}''\cite{Ferziger1966}).
Nowadays, there are much more sophisticated and advanced theories, each working
more or less satisfactorily in its area, but the practical considerations most
often prevail~\cite{MacFarlane2010,Kok2016}.

In the absence of a convenient theory of neutron moderation, another
possibility to deal with complex geometries, different medium compositions, and
temperatures, is to use some Monte Carlo simulations which allow to tune all
these parameters and see how they influence on the shape of the resulting 
steady-state neutron spectrum.

%

\begin{figure}[tb!]
\begin{center}
\includegraphics[width=12cm]{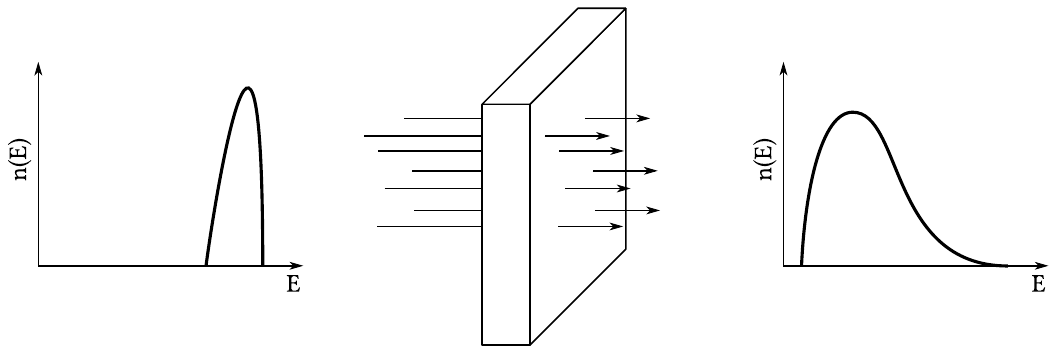} \\
\includegraphics[width=12cm]{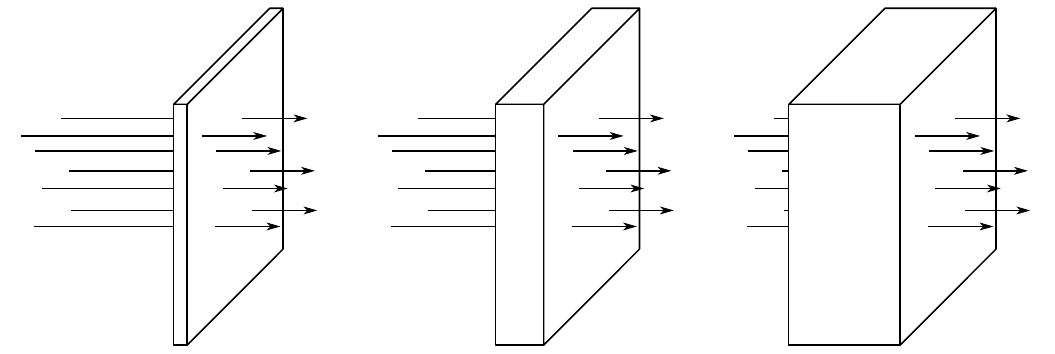}
\end{center}
\caption{A sketch of the approach to the study of neutron spectrum moderation
when the initially high-energy neutrons (upper left panel) is passed through
some layer of moderator and shifts to the low energies (upper right panel).
Changing the moderator layer thickness, it is possible to make some conclusions
on dynamics of neutron moderation (bottom panel).}
\label{fig-neutrons-through-layer}
\end{figure}

In Monte Carlo simulations of the neutron moderation by some medium the
following approach is often used. The initial neutron spectrum is set, and the
neutrons are directed to some moderating layers (optionally of variable
thickness). The passage of neutrons through this medium is tracked, and the
spectrum of neutrons coming at the opposite side is measured
(Fig.~\ref{fig-neutrons-through-layer}). By comparing the
resultant spectrum to the initial one for different moderator thickness, some
conclusions can be made on how the moderator layers influence the initial
neutron spectrum~\cite{Deiev2013,Lisovska2019,Shin2014,Robinson2014,Lemrani2006}.

However, such approach is rather limited, and is disadvantageous for our task. 
The problem is that the Monte Carlo codes such as Geant4 treat the injection of each primary particle as a separate \textit{Event}, and simulate its further interactions independently of all the other primary particles.
It means that each primary particle is injected into the system at "zero" time (\textit{i.e.} the particle enters a "fresh" system with no signs of other particles' presence or previous interactions), and is then tracked down to its disappearance either by absorption or by simply escaping the volume of interest. This is of course accompanied by the tracking of all the secondary particles if there are some.


Repeating this scenario $N$ times for a large number of primary particles is actually equivalent to injecting a \textit{single pulse} of $N$ primary particles into the system.
And the spectrum thus obtained is therefore a spectrum of a
\textit{single} ``generation'' of neutrons which were born (injected) at the
same time, then passed through some moderating layers, and eventually formed
some final distribution at the output.

But how would the form of the final spectrum change if we used a fixed neutron source, and the new particles
were born and added to the system all the time?
The new (``younger'') particles
would obviously possess their initial energy, and they would be added to the
``older'' particles which had already slowed down to lower energies. The total
spectrum of such system will strongly depend on the time that the neutrons possess a
certain energy (or the time the neutrons spend in a certain energy bin), as
this determines the appearance of the spectral maxima and minima. For example,
the maximum is expected to appear in the energy range in which the neutrons
spend most of their time, and the fixed neutron source will make them
gradually accumulate there (older ones + younger ones + even younger ones etc.).

For this reason, in problems involving the action of a fixed neutron source, 
it is necessary to work out some other way of calculating the neutron spectrum in Geant4 with due account taken of \textit{time}.

In this paper we focus on the problem of
predicting the steady-state neutron spectrum which sets in some medium under
the action of a fixed neutron source, using the 
philosophy
and capabilities of
the Geant4 library~\cite[Chapter V. Tracking and physics]{GEANT4DevBook2022}, and restricting the consideration to the subcritical cases
only (simulation of the super-critical conditions in Geant4 requires some
special treatment which is not the subject of the present paper).
In particular, in Sections~\ref{sec-timeslices-idea}-\ref{sec-timeslices-Geant4} we describe
our approach to the steady-state neutron spectrum calculation in Geant4.
In Section~\ref{sec-hydrogen-test} we perform the basic test of our
algorithm for the simplest case of neutron moderation in pure hydrogen. Next,
in Section~\ref{sec-logscale} we discuss the advantages of the use of the time
slices with the logarithmic temporal step, and describe a method of their
implementation in Geant4. In Section~\ref{sec-no-timeslice} we also describe an
alternative way of building the steady-state neutron spectrum without the time
slices, and discuss its pros and cons. Finally, in 
Section~\ref{sec-optimization} we discuss some technical details of the
possible simulation optimizations.

\section{The concept of time slices of neutron spectrum}
\label{sec-timeslices-idea}

In the framework of Monte-Carlo method it is possible to track the change in
energy of a single neutron with time in certain medium. This procedure may be
repeated many times from the start, imitating a simultaneous injection of a
large number of neutrons into the studied system. This also allows to set the
initial neutron spectrum of any form.

Let us conventionally call such a single portion of neutrons simultaneously
injected into the system, a single ``generation of neutrons''.

\begin{figure}[tb!]
\begin{center}
\includegraphics[width=10cm]{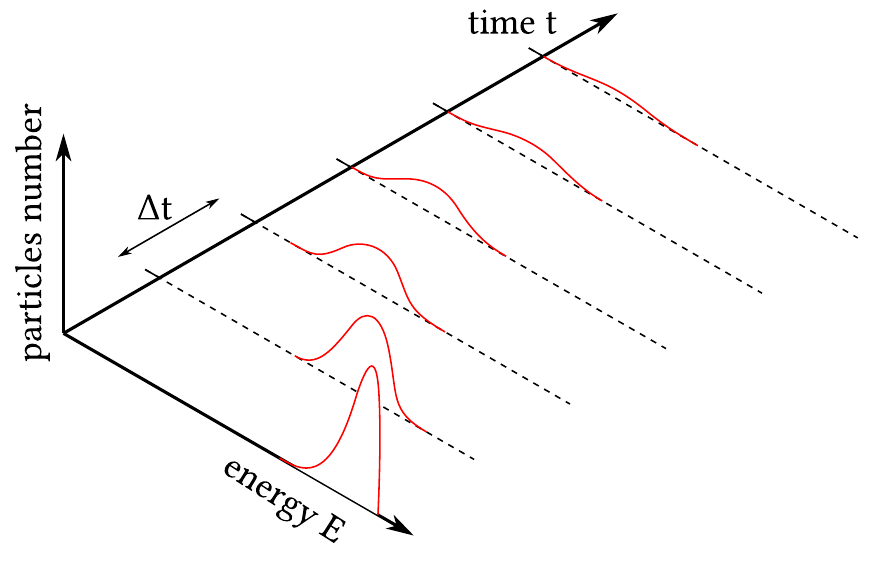}
\end{center}
\caption{Time slices of some neutron spectrum taken every $\Delta t$ seconds.
This figure represents the case when the initially high-energy spectrum
gradually shifts to the low energy range with time.}
\label{fig-slices-linear}
\end{figure}

If we know the neutron energies at any time, it is possible to build the
``snapshots'' of their spectra at different moments in time. We shall call them
the \textit{time slices} of the neutron spectrum. Such time slices will show
the temporal evolution of the neutron spectrum in some medium with chosen
parameters (Fig.~\ref{fig-slices-linear}).

Within the framework of Monte-Carlo method, it is reasonable to assume that the
following neutron generations will demonstrate a similar spectrum evolution,
provided that all of them are produced under similar conditions. This means
that starting with zero time, after a period of $\Delta t$, the spectrum of the
first generation of neutrons will be $n(E,\Delta t)$. During this period a
second generation of neutrons will be injected into the system from the neutron
source with initial spectrum $n(E,t=0)$. After another period $\Delta t$ the
first generation of neutrons will be moderated to $n(E,2\Delta t)$, and the
second one -- to $n(E,\Delta t)$, and also the third generation of neutrons
will be injected with their initial spectrum. And so on and so forth.

After some time $T_{max}$, when the first generation of neutrons is totally
thermalized and absorbed (or escapes from the system), a stationary state will
settle -- each leaving generation of neutrons will be replaced by a younger one.
The total neutron spectrum will stop changing, and may be considered a
\textit{non-equilibrium stationary spectrum} for the considered combination of
the chosen medium and neutron source. This is of course true for the case of a
constant neutron source only.

To determine such non-equilibrium stationary neutron spectrum is the main goal
of the present research.

So in the volume which contains all of the mentioned neutron generations at
once, the total spectrum is composed of the spectra of all these generations:

\begin{equation}
n(E) = \int \limits _{0} ^{T_{max}} n(E,t) dt ,
\label{eq02}
\end{equation}

\noindent
where $T_{max}$ is the neutron lifetime in the studied system, $n(E,t)$ is the
spectrum of neutrons which spent the time $t$ in the system.
Or in the discrete case:

\begin{equation}
n(E) = \sum \limits _{i=0} ^N n_i (E) \cdot \Delta t ,
\label{eq03}
\end{equation}

\noindent
where $n_i (E)$ is the spectrum of neutrons in $i^{th}$ time slice, $\Delta t$
is the time bin width.

The time bins are not necessarily of equal width, so in general case,

\begin{equation}
n(E) = \sum \limits _{i=0} ^N n_i(E) \cdot (\Delta t)_i ,
\label{eq04}
\end{equation}

\noindent
where $(\Delta t)_i$ is the width of the $i^{th}$ time bin.

So, in order to apply all these considerations to some Monte Carlo simulation,
it is necessary to find a way to calculate the mentioned spectrum time slices
at chosen moments in time.

\section{Time slices in Geant4}
\label{sec-timeslices-Geant4}

The Geant4 Monte-Carlo engine~\cite{Geant4Toolkit2003,Geant4applications2006,
Allison2016} generates the primary particles only one at a time. The particle's
trajectory is being tracked from the moment of its appearance in the system --
till the moment of its vanishing or complete stop~\cite{GEANT4DevBook2022}.
After that another primary particle is generated and its trajectory is tracked.
The trajectory thus obtained is a polygonal line consisting of consecutive
steps. Each step is limited by its \texttt{PreStepPoint} and
\texttt{PostStepPoint}.

Each \texttt{StepPoint} stores the information on:
\begin{itemize}
\item particle position
\item particle momentum direction
\item particle kinetic energy
\item global time
\item and some other useful quantities
\end{itemize}

Since each step is generated randomly, it is impossible to predict (or preset)
the moments $t$ at which the particle appears in \texttt{PreStepPoint} or
\texttt{PostStepPoint} (these points are chosen randomly according to the
physical processes applied). Therefore it is also impossible to obtain the
neutron spectrum ``snapshot'' at arbitrary moment in time, as was described in
Section~\ref{sec-timeslices-idea}. However, due to the fact that Geant4 stores
the \texttt{GlobalTime} for each \texttt{StepPoint} of the particle track, it
is possible to overcome this limitation.

\begin{figure}[tbp!]
\begin{center}
\includegraphics[width=14cm]{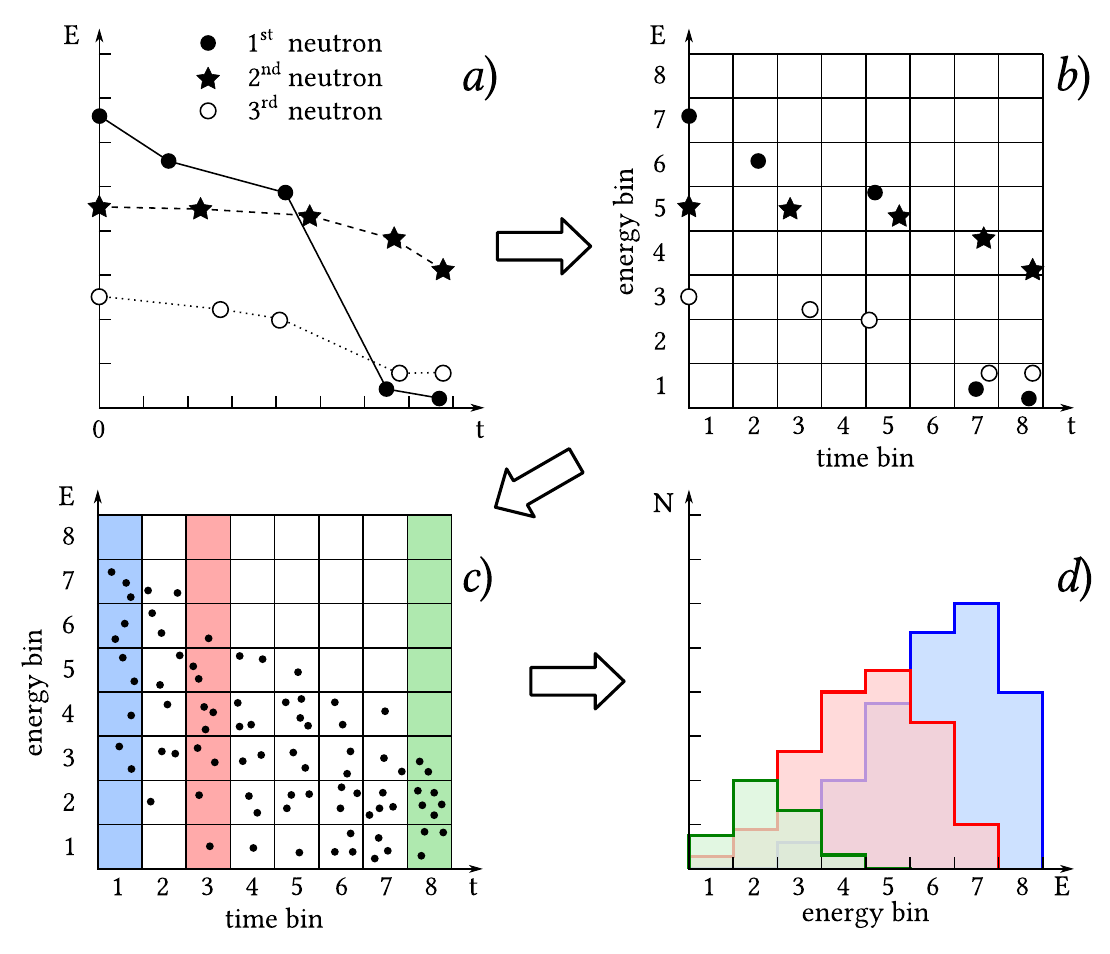}
\end{center}
\caption{A method of time slices retrieval for the neutron spectrum in Geant4:
a) first, the life of each neutron is simulated separately;
b) all the registered neutron appearances are associated with the corresponding
energy and time bins;
c) the accumulation of (very) large number of events in time bins;
d) plotting of the accumulated spectrum snapshots and summing them up to obtain
a total stationary neutron spectrum which settles in the system.}
\label{fig-spectra-accumulation-Geant4}
\end{figure}

One can divide the time axis into intervals -- time bins -- and using the
values of energy and \texttt{GlobalTime} at each \texttt{StepPoint}, put the
particle into the corresponding time bin (Fig.~\ref{fig-spectra-accumulation-Geant4}).

For the algorithm sketched in Fig.~\ref{fig-spectra-accumulation-Geant4} a
simple 2D array may be used. The dimensions of this array are the number of
bins in the energy spectrum  -- \texttt{numEnergyBins} (often referred to as
``channels'' in spectrometry), and the number of time intervals we divided the
time scale into  -- \texttt{numTimeBins}.

The \texttt{GlobalTime} of the current \texttt{StepPoint} as well as the
current kinetic energy of the tracked particle may be accessed in the
\texttt{UserSteppingAction} function of the \texttt{G4UserSteppingAction}
class. The corresponding time and energy bins may be calculated as follows:

\begin{sourcecode}{}
G4double currentTime = step->GetPreStepPoint()->GetGlobalTime(); 
G4double currentKinE = step->GetPreStepPoint()->GetKineticEnergy();

G4int energyBin = floor((currentKinE)/(maxEnergy/numEnergyBins));
G4int timeBin   = floor((currentTime)/(maxTime/numTimeBins));
\end{sourcecode}

Let us consider all possible cases of neutron registration and distribution to
the corresponding bins (Fig.~\ref{fig-neutrons-in-bins}):

\begin{figure}[tbp!]
\begin{center}
\includegraphics[width=14cm]{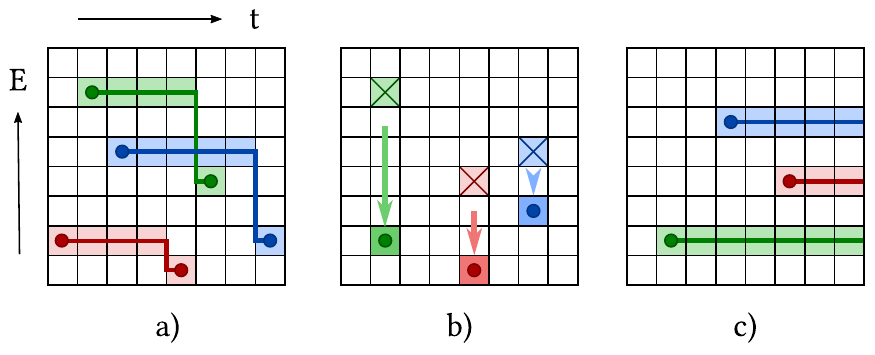}
\end{center}
\caption{a) Neutrons with different energies in different time bins.
b) Neutrons in the same time bin -- we keep the last entrance only.
c) Neutrons born at later times and falling into the time bins other than
zero.}
\label{fig-neutrons-in-bins}
\end{figure}

\begin{enumerate}

\item The neutron is registered for the first time with zero
\texttt{GlobalTime}. In this case it is considered to be the initial neutron,
and its energy lets us to fill in the corresponding cell in the 2D 
time-spectrum array.

\item The neutron is registered once again with different energy and at
different time. In this case we assume its energy did not change during all the
intermediate time bins, and we fill in those array cells as well
(Fig.~\ref{fig-neutrons-in-bins}a).


\begin{sourcecode}{}
if (!isFirstStep)
  for(int i=lastTimeBin; i<timeBin; i++)
    pTimeSpectrum[lastEnergyBin][i] += 1;
\end{sourcecode}

Here \texttt{isFirstStep} is \texttt{true} for the first fire of a neutron,
\texttt{lastTimeBin} is the previous time bin, in which the neutron had been
registered earlier, \texttt{timeBin} is the current time bin, in which the
neutron is registered for the second time, \texttt{lastEnergyBin} is the
previous energy bin, in which the neutron had been registered, the
\texttt{numTimeBins} is the total number of time bins, and the
\texttt{pTimeSpectrum} is the pointer to the 2D array containing the spectra.

\item The neutron hits the same time bin for the second time, but with
different energy. In this case we have to subtract it from the previous energy
bin and add it to a new energy bin (Fig.~\ref{fig-neutrons-in-bins}b).

\begin{sourcecode}{}
pTimespectrum[energyBin][timeBin] += 1;
if ((timeBin == lastTimeBin) && (!isFirstStep))
  pTimeSpectrum[lastEnergyBin]][timeBin] -= 1;
\end{sourcecode}

\item The neutron is registered for the first time with non-zero
\texttt{GlobalTime}. Such neutron is considered a secondary particle, so there
is no need to fill in all the earlier time bins
(Fig.~\ref{fig-neutrons-in-bins}c). This was actually the reason to check the
\texttt{isFirstStep} value in the 2$^{nd}$ point above.

\end{enumerate}

\section{Hydrogen test}
\label{sec-hydrogen-test}

As the analytical neutron moderation theory is best established for the
hydrogen nuclei (namely, the protium), it seems reasonable to start the
verification of the algorithm from this material.

Any simulation in Geant4 starts with the construction of the system's geometry and definition of all the substances and materials used. Here we use the hydrogen density at STP:

\begin{sourcecode}{}
G4NistManager* nistMan = G4NistManager::Instance();
G4double density = 8.988e-2 * g/cm3;
G4double temperature = 273.15 * kelvin;
G4double pressure = 1e5 * pascal;
G4Material *Hydrogen = 
nistMan->BuildMaterialWithNewDensity("Moderator", "G4_H", 
                                     density,
                                     temperature,
                                     pressure);
\end{sourcecode}

By this test, the neutron moderation theory~\cite{Tarasov2017Moderation} taking
into account the medium temperature may also be supplemented with the Monte
Carlo simulations.  
For this purpose we are currently interested
in simulating the neutron moderation in an infinite homogeneous medium first.

To simulate an infinite medium, we use a ``\texttt{Detecor}'' in a form of a
cube with the side comparable to the size of the visible Universe
($\sim 10^{26}m$).

\begin{sourcecode}{}
G4Box *world_box = new G4Box("world", 1e26*m, 1e26*m, 1e26*m);
\end{sourcecode}

All this cube will act as a moderator and a sensitive volume at the same time. It means that the entire cube is filled uniformly with hydrogen atoms, and the neutrons are being tracked throughout its entire volume.

Next it is necessary to define the physical processes. We choose the following set of processes described in Geant4 library:
\begin{itemize}
\item electromagnetic interactions \cite{Allison2016},

\item elastic hadron interactions \cite{GEANT4PhysRef2017}, 

\item QGSP\_BERT\_HP (``Quark-gluon String, Precompound, Bertini, High
Precision'') which uses the quark-gluon model for the hadron energies above
12~GeV \cite{Apostolakis2009,Wright2015} (such energies are not required for
our current purpose, but makes the program more flexible and adjustable to
future applications), and the high-precision data on neutron scattering,
capture and fission from the G4NDL database \cite{G4NDL}.
\end{itemize}

\begin{sourcecode}{}
#include "G4EmStandardPhysics.hh" 
#include "G4HadronElasticPhysicsHP.hh" 
#include "G4HadronPhysicsQGSP_BERT_HP.hh" 
PhysicsList::PhysicsList() {
    RegisterPhysics(new G4EmStandardPhysics()); 
    RegisterPhysics(new G4HadronElasticPhysicsHP(0)); 
    RegisterPhysics(new G4HadronPhysicsQGSP_BERT_HP(0)); 
}
\end{sourcecode}

Finally, a \texttt{PrimaryGeneratorAction} class must be defined. This class
describes the source of initial particles. For our test it may be
a simple point source located at the center of the ``\texttt{Detector}'' cube:

\begin{sourcecode}{}
PrimaryGeneratorAction::PrimaryGeneratorAction() {
    particleGun = new G4ParticleGun(1);
    particleGun->SetParticleDefinition(G4Neutron::Neutron());
}

void PrimaryGeneratorAction::GeneratePrimaries(G4Event* event) {
    particleGun->SetParticlePosition(G4ThreeVector(0, 0, 0));
    particleGun->SetParticleMomentumDirection(G4RandomDirection());
    particleGun->SetParticleEnergy(2*MeV); ///constant energy
    particleGun->GeneratePrimaryVertex(event);
}
\end{sourcecode}

Note that under such conditions (an extremely large moderator volume and the neutron source located at its center), there is no realistic way for a neutron to escape from the moderator. Therefore, the only possible way for a neutron to disappear from the system (and stop being tracked by Geant4 engine) is to be absorbed by the moderator nuclei.

\begin{figure}[tbp!]
\begin{center}
\includegraphics[width=16cm]{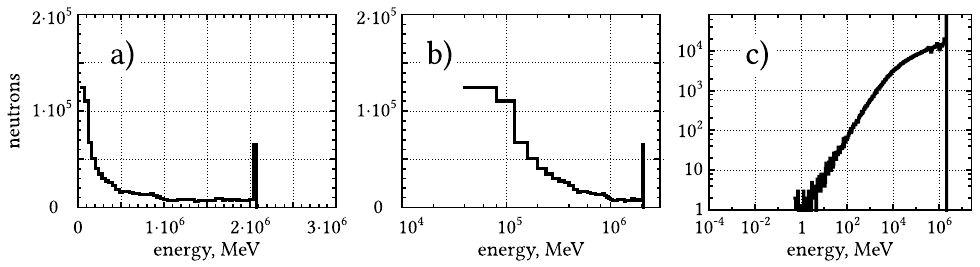}
\end{center}
\caption{Neutron moderation spectrum in water for the initial energy of
$\sim 2~MeV$.
\textbf{a)} Both energy scale and neutron count scale are linear.
\textbf{b)} Linear energy bins during calculation depicted in the logarithmic
scale.
\textbf{c)} Logarithmic energy bins during calculation.}
\label{fig-different-energy-scales}
\end{figure}

An example spectrum of neutrons obtained this way is shown in 
Fig.~\mbox{\ref{fig-different-energy-scales}} in three different forms. As can be seen from this figure, the
most convenient visual form of the spectra is the log-log scale
(Fig.~\ref{fig-different-energy-scales}c), which requires the use of the
logarithmic energy bins at the stage of calculations. Such scale will be used
in most cases below.

The corresponding energy bin should be calculated in the following way:

\begin{sourcecode}{}
G4int energyBin =
  floor((log10(currentKinE/MeV) - minOrder) /
        ((log10(maxEnergy/MeV)  - minOrder) / numEnergyBins));
\end{sourcecode}

\noindent 
where \texttt{numEnergyBins} is the number of energy bins, \texttt{currentKinE}
is the current kinetic energy, \texttt{minOrder} is the minimum required
magnitude order for the energy, \texttt{maxEnergy} is the maximum possible
energy. This way we obtain a form of the spectrum as in
Fig.~\ref{fig-different-energy-scales}c.

In the context of nuclear reactor applications, a Watt fission spectrum may be
used for the initial neutrons:
\begin{equation}
\rho(E) = c \cdot \exp{(-a E)} \cdot \sinh{(\sqrt{b E})} ,
\end{equation}
\noindent where $a = 1.036$, $b = 2.29$ and $c = 0.4527$ for $^{235}U$ fission
spectrum.

\begin{figure}[tbp!]
\begin{center}
\includegraphics[width=16cm]{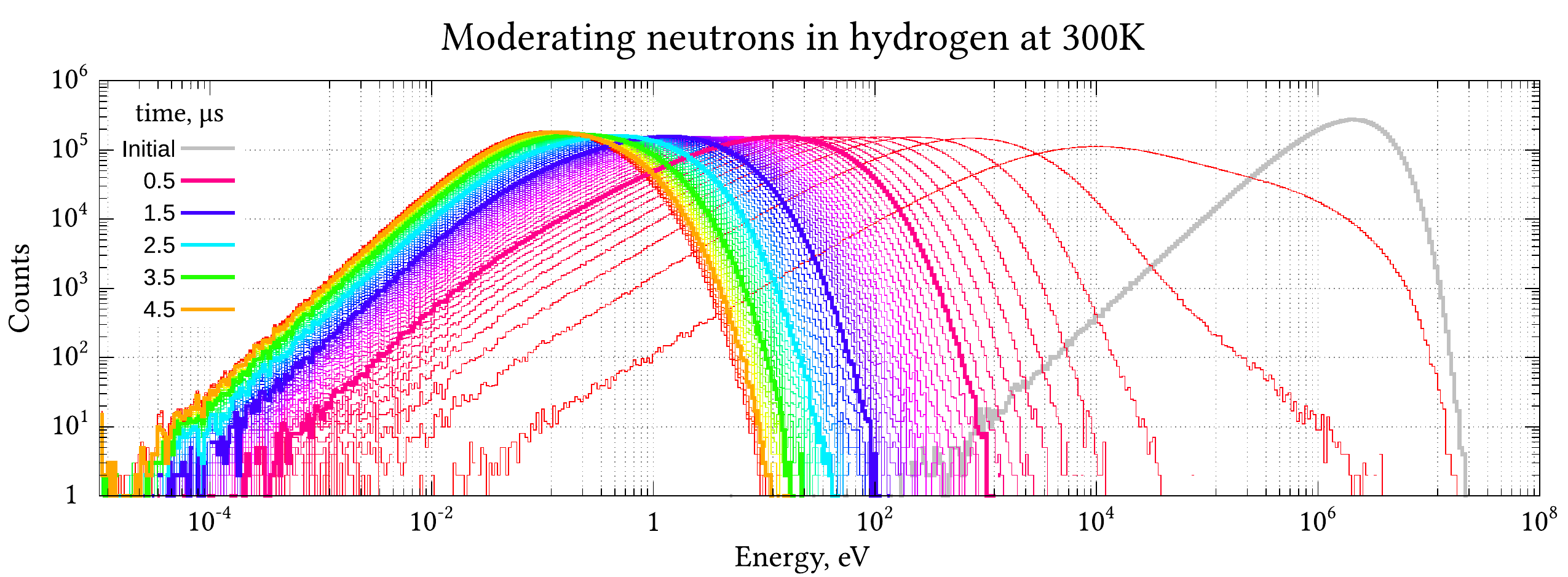}
\end{center}
\caption{Simulated evolution of the neutron spectrum in hydrogen at 300K. A
fission spectrum was used initially (grey line). The colored lines show the
transitional states of the neutron spectrum up to 4.5~$\mu$s. The spectrum
evolution is very rapid at first, and becomes more graceful later, which is
why the time slices are not evenly spaced, although taken in equal intervals.}
\label{fig-spectrum-evolution-lintime}
\end{figure}

Now everything is ready to run the simulation. We generate a sufficiently large number of primary neutrons whose initial energies are distributed according to a fission spectrum of $^{235}U$, given above, and then apply the algorithm described in Section~\mbox{\ref{sec-timeslices-Geant4}}, to each of them. Finally, we get the time slices of the neutron spectrum of a single generation of neutrons (remember, all the primary neutrons were generated under exactly the same conditions and tracked independently, which is equivalent to injecting them into the moderator at zero time, as was discussed above). These time slices are shown in Fig.~\mbox{\ref{fig-spectrum-evolution-lintime}}.

\begin{figure}[tb!]
\begin{center}
\includegraphics[width=16cm]{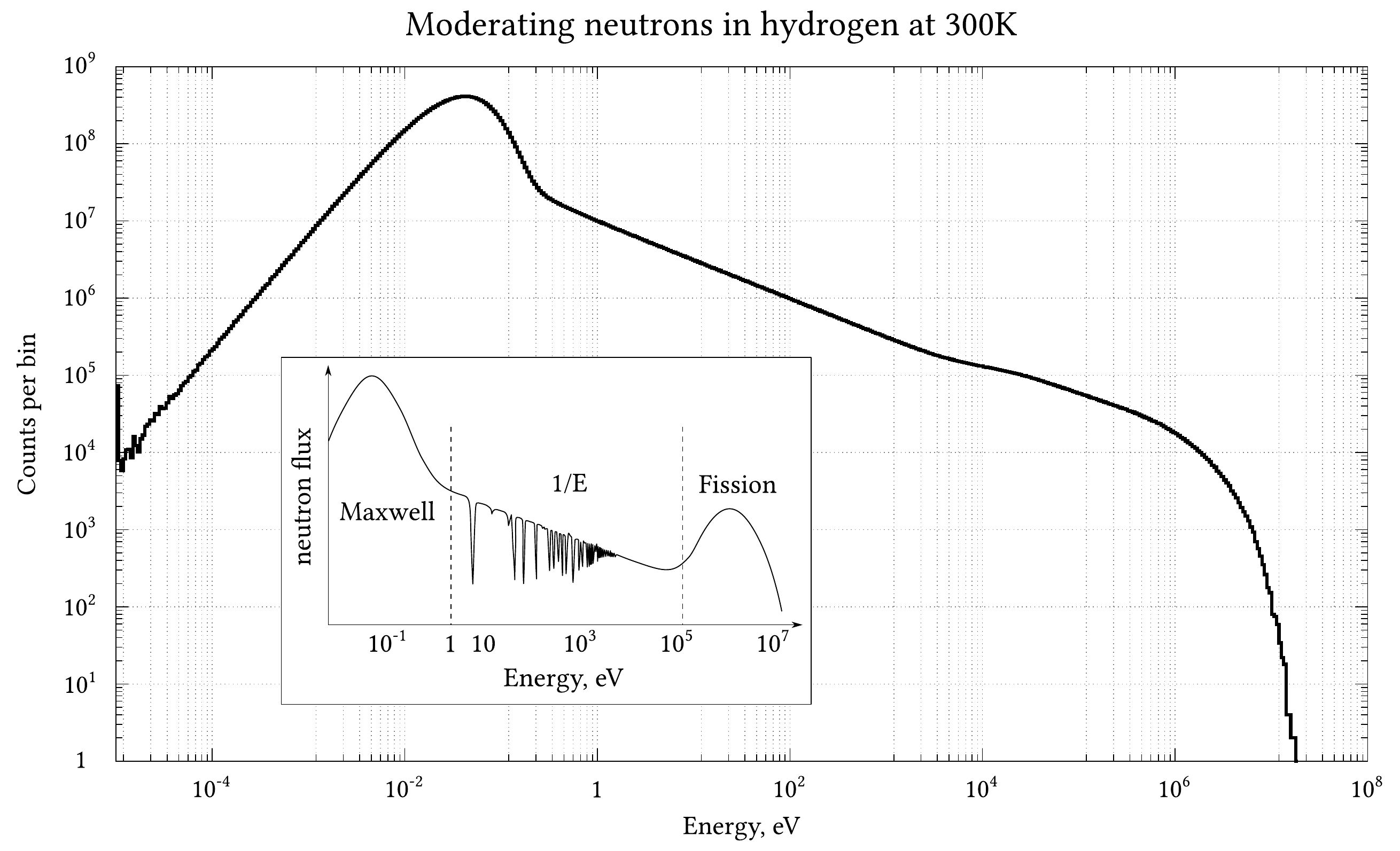}
\end{center}
\caption{Total stationary spectrum of neutrons in hydrogen calculated by
summing up all the contributions from the transitional neutron spectra shown
in Fig.~\ref{fig-spectrum-evolution-lintime} up to 100~$\mu$s. Inset: the form
of a neutron spectrum in thermal reactor from Fig.~2.1 in~\cite{Oka2014}.
Note that the main image shows the neutron population (not flux) distribution
over the logarithmic bins (bin width is proportional to energy), so the $1/E$
slope is consistent between the main image and the inset.}
\label{fig-total-spectrum-lintime}
\end{figure}

Figure~\mbox{\ref{fig-total-spectrum-lintime}} shows the respective stationary neutron spectrum calculated from all the intermediate ones (the time slices) using Eq.~(\mbox{\ref{eq03}}).
As can be seen in 
this figure,
the stationary neutron
spectrum in hydrogen moderator has exactly the expected form. 
Indeed,
the
comparison with the typical neutron spectrum in thermal reactor (inset in 
Fig.~\ref{fig-total-spectrum-lintime}) shows that there is an extremely close
correspondence, aside, of course, from the fact that our spectrum was simulated
in the pure hydrogen medium, which explains the absence of the resonance area and
a substantially lower level of fission spectrum.

This basic test of the neutron moderation in pure hydrogen, although may seem oversimplified, is still very important, as it validates our approach and proves that the algorithm described in Section~\mbox{\ref{sec-timeslices-Geant4}} is able to produce correct results. Since this algorithm is rather generic, and does not impose any restrictions on the moderator's geometry and material, as well as on the form of the primary particle source and initial energy, it can be easily applied to arbitrary medium of any form and composition, and to any source of primary particles.

For example, in order to change the geometry of the studied medium, instead of using a simple cube (\texttt{G4Box} class) like we did in our test above, one can create instances of any classes describing solids -- either the basic shapes (e.g. \texttt{G4Tubs}, \texttt{G4Cons}, \texttt{G4Sphere}, and so on), or even more complex tesselated solids. Geant4 allows also to put one solid into another (hierarchical structure), or produce the unions, intersections or subtractions of solid volumes~\mbox{\cite[Section~4.1.2 Solids]{GEANT4DevBook2022}}.

In our test we used a standard hydrogen from the NIST materials database (code \texttt{G4\_H}). However, each of the created volumes may have its own material (described by the \texttt{G4Material} class) including its state, temperature, density, pressure, and of course, the isotope composition, which is very important in our case.
In fact, the simulated medium may even include some fissile materials (acting as neutron multiplicators), as long as the sub-criticality condition is met. In this case, the multiplication of neutrons due to fission reactions will be treated correctly by the Geant4 engine, since the number of neutrons always remains finite (due to sub-criticality condition). A very handy \texttt{G4NistManager} class gives quick access to a rich collection of predefined widely used materials.

In order to change the parameters of the primary particle source (the particle type, its initial energy and momentum, and the position of its first appearance in the simulated volume, one can either set them for the corresponding \texttt{G4ParticleGun} instance while constructing a necessary \texttt{G4VUserPrimaryGeneratorAction}-derived class, or even use a universal \texttt{G4GeneralParticleSource} class to allow the user to load those parameters from an external macro file at run-time without the need of re-compilation.

Such flexibility of Geant4 allows us to apply our algorithm described in Section~\mbox{\ref{sec-timeslices-Geant4}}, to a variety of tasks beyond reactor physics, ranging from e.g. the radiation shielding and protection to medical imaging applications and radiation therapy.

\begin{figure}[tbp!]
\begin{center}
\includegraphics[width=8cm]{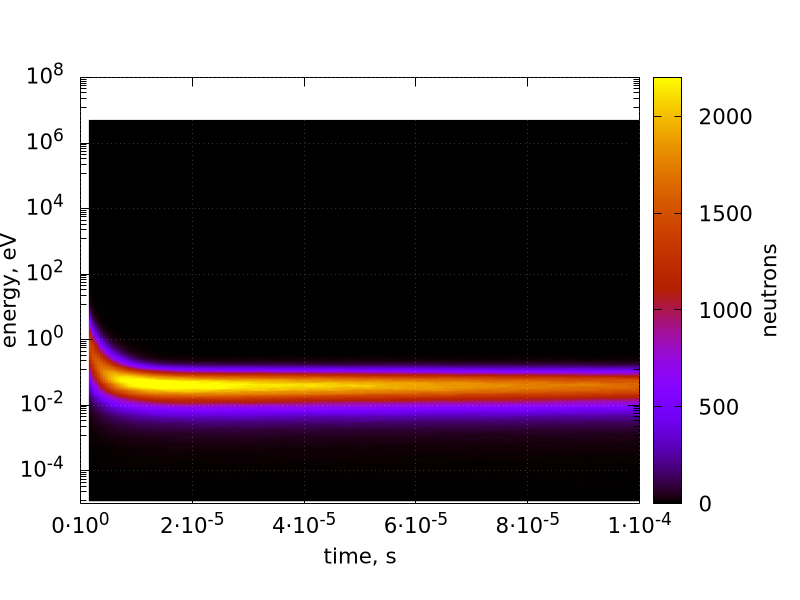}
\hfill
\includegraphics[width=8cm]{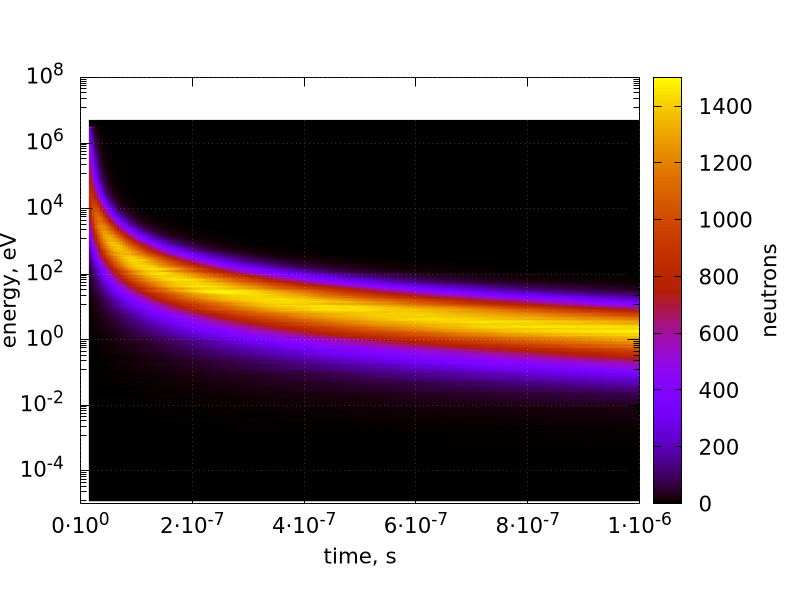}
\end{center}
\caption{Pseudo-3D view of the neutron spectrum evolution. The time increases
along the X-axis, the energy bins are along the Y-axis, and the color scale
represents the number of neutrons in corresponding bins.}
\label{fig-pseudo3D-evolution}
\end{figure}

Turning back to the analysis of the results of our hydrogen test, it is interesting to note the evolution of the spectrum shape with time, shown in Fig.~\mbox{\ref{fig-spectrum-evolution-lintime}} (please note that the spectra shown in this figure represent the ``snapshots'' of the spectrum of a \textit{single generation} of neutrons, or alternatively, a \textit{single pulse} of neutrons, taken at different times).
For a different visual representation of the spectrum evolution it is 
possible to build a kind of pseudo-3D (or even a 3D) plot 
(Fig.~\ref{fig-pseudo3D-evolution}).
As can be seen in Fig.~\mbox{\ref{fig-pseudo3D-evolution}}, the distribution of neutron energies shifts to the thermal area very quickly, and then its position remains the same for quite a long time, and it just gradually fades out as the neutrons get absorbed by the medium. This picture served as a motivation to the use of a logarithmic time scale described in the next section.

\FloatBarrier

\section{Logarithmic time scale}
\label{sec-logscale}

Fig.~\ref{fig-pseudo3D-evolution} may serve as an example of the case when the
neutrons are moderated very rapidly at first, and then live for quite a long
time in the thermal area with little change in their spectrum. It is
hard to see any details of the spectrum
evolution in such view. Therefore, by analogy to what we have done with the
energy distribution earlier, we can apply the logarithmic scale for the time
axis here. The basic idea is to make the earlier time bins short, and the later
ones -- long. As long as the spectrum changes noticeably at first stages only,
and the later time slices do not differ drastically from each other, this
procedure will not introduce any serious error.

The use of the logarithmic time scale has also some technical advantages. In
order to catch the details of rapidly changing neutron spectrum at early
stages, the time bin width $\Delta t$ must be chosen rather small. On the other
hand, in order to track the neutrons till the very end of their lifetime in a
weakly-absorbing medium, the maximum simulated time $T_{max}$ must be chosen
rather large. Putting these two conditions together leads to a large overall
number of time bins required. It can make the 2D \texttt{TimeSpectrum} array
too large and inconvenient to operate.

\begin{figure}[tbp!]
\begin{center}
\includegraphics[width=10cm]{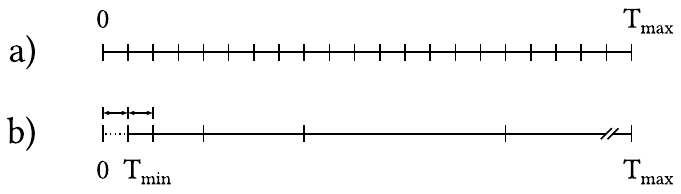}
\end{center}
\caption{a) Linear scale for time bins. b) Logarithmic scale for time bins.
The first and second time bins are of equal length, so the actual logarithm is
applied starting from $T_{min}$ only, and each next time bin is twice as large
as the previous one.}
\label{fig-log-timebins-scheme}
\end{figure}

In case of the energy scale it was not necessary to start with zero, so it was
possible to choose some minimum energy and simply calculate the logarithm from
there on. As to the time scale, it must start from zero, so applying a
logarithm requires some special care. Here we suggest to choose the maximum
time $T_{max}$ and the desired number of time bins $m$, and then by imposing a
condition that the first and the second bins must be of equal width, find the
start of the logarithmic time scale $T_{min}$ (see 
Fig.~\ref{fig-log-timebins-scheme}) as follows:

\begin{equation}
T_{min} = \frac{T_{max}}{2^{(m - 1)}}
\label{eq-Tmin}
\end{equation}

Given the current time $t$, the corresponding time bin may be calculated as
follows:

\begin{equation}
bin = \left \lfloor \frac{(m-1) \cdot \log_{10} (t / T_{min})}{\log_{10}(T_{max} / T_{min})} +1 \right \rfloor
\end{equation}

For the times less than $T_{min}$ the logarithm is negative, so all such values
should go into the $1^{st}$ time bin (number zero in terms of C/C++ syntax).

In fact, 2 in Eq.~(\ref{eq-Tmin}) represents the ratio of the two adjacent bins,
so in general case, it may be substituted by some other desirable value depending on the task.

\begin{equation}
T_{min} = \frac{T_{max}}{binRatio^{(m - 1)}}
\end{equation}

Since the time bins are different in length here, the total stationary spectrum
should be calculated using Eq.~(\ref{eq04}).

\begin{figure}[tbp!]
\begin{center}
\includegraphics[width=16cm]{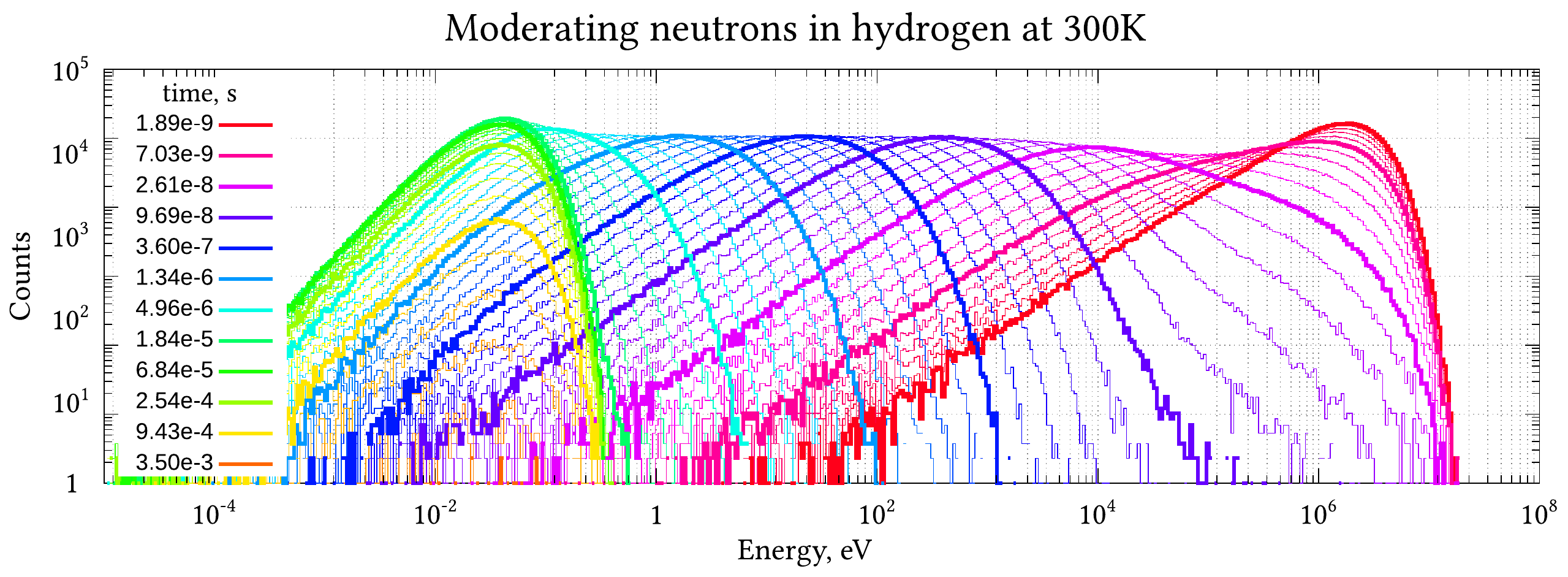}
\end{center}
\caption{Time evolution of neutron spectrum in hydrogen, built using the
logarithmic time bins. The logarithmic time scale is apparently better suited
for depicting the neutron spectrum evolution, as it allows to reflect the 
detailed form of the spectrum at both very short and very long time intervals.}
\label{fig-spectrum-evolution-logtime}
\end{figure}

\begin{figure}[tbp!]
\begin{center}
\includegraphics[width=8cm]{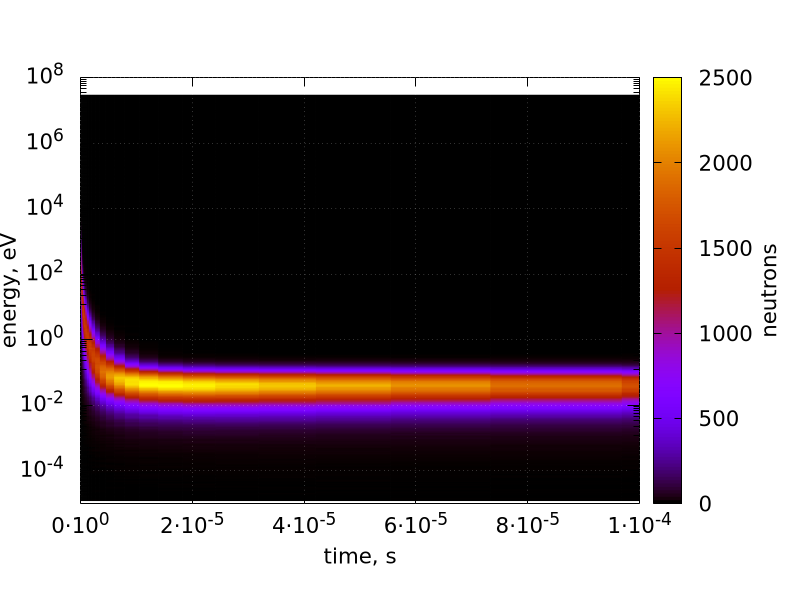}
\includegraphics[width=8cm]{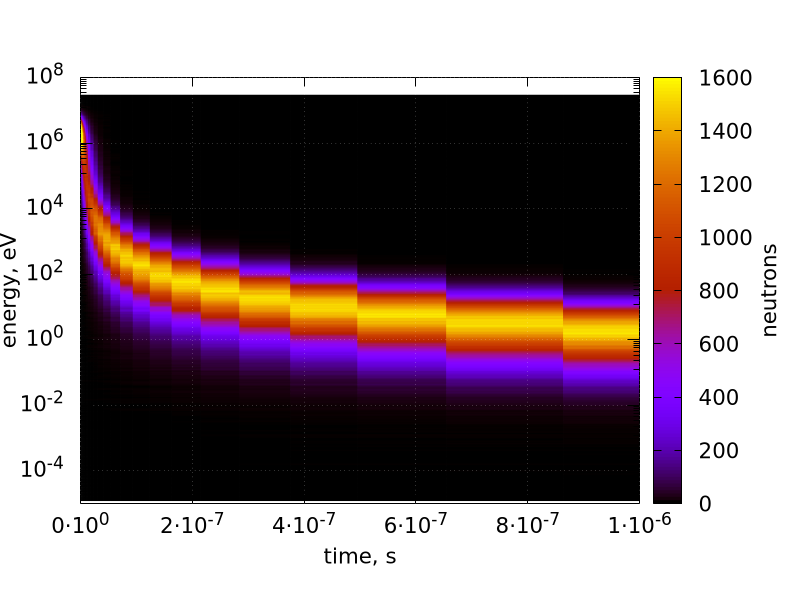}
\end{center}
\caption{Pseudo-3D images of the spectrum evolution in Hydrogen simulated using
the logarithmic time bins. Compare to the Fig.~\ref{fig-pseudo3D-evolution}.
Note that here we used only 75 time bins for the left panel (up to $100~\mu s$)
and 67 bins for the right panel (up to $1~\mu s$) as opposed to the 2000 time
bins for both cases in Fig.~\ref{fig-pseudo3D-evolution}.}
\label{fig-pseudo3D-evolution-logtime}
\end{figure}

The time slices of neutron spectrum in hydrogen taken at logarithmic time steps
$(\Delta t)_i$ are shown in Fig.~\ref{fig-spectrum-evolution-logtime} as 2D
time slices, and in Fig.~\ref{fig-pseudo3D-evolution-logtime} as pseudo-3D
graphs. 
These images, and especially Fig.~\ref{fig-spectrum-evolution-logtime}, demonstrate that, in addition to the calculation of a steady-state spectrum, the suggested approach gives us access to the spectrum shapes at different moments in time (or rather, at different stages of moderation) in a wide range, starting from the particle entrance to the moderator medium, and on. It could be an interesting task to analyze the statistical parameters of those distributions to study the process of neutron moderation in more detail. The authors have made some investigations of the spectrum variance and entropy for several materials in~\cite{Smolyar2021JPS}.

\begin{figure}[tbp!]
\begin{center}
\includegraphics[width=16cm]{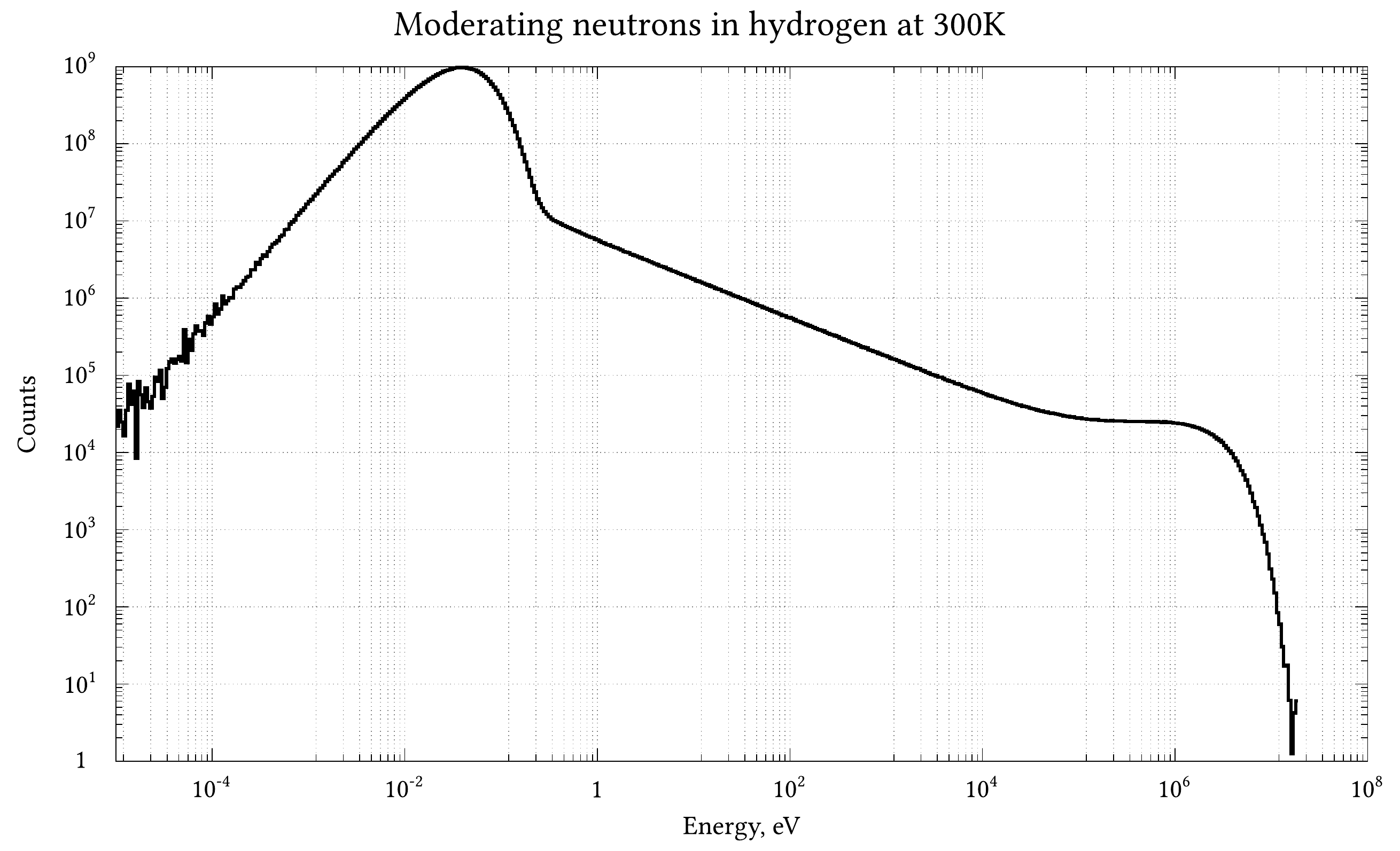}
\end{center}
\caption{Total stationary spectrum of neutrons in hydrogen at 300K calculated
by summing up all the contributions from the transitional neutron spectra shown
in Fig.~\ref{fig-spectrum-evolution-logtime} up to $10^{-2}$~s.}
\label{fig-total-spectrum-Hydrogen-logtime}
\end{figure}

The total stationary neutron spectrum calculated on the basis
of the time slices thus taken is shown in 
Fig.~\ref{fig-total-spectrum-Hydrogen-logtime}. It is apparently somewhat more
accurate than the one in Fig.~\ref{fig-total-spectrum-lintime} because of
a) the more detailed account for the spectrum evolution at early stages, and
b) the larger time scales available (10~ms versus 100~$\mu$s in 
Fig.~\ref{fig-total-spectrum-lintime}).

One remark should be made here. For certain media, a production of secondary
neutrons is possible (the case of neutron multiplying media mentioned earlier). Such neutrons appear for the first time in some later
time bins (corresponding to non-zero time). Since the later time bins are
wider than the earlier ones, the contribution of such neutrons may be
overestimated when multiplied by the corresponding bin width $(\Delta t)_i$.
One option to overcome this drawback may be to use a 2D array of real numbers
instead of integers, and instead of incrementing its values by a unit per each
neutron, increment them by a fraction of time bin during which the neutron
actually existed.

\FloatBarrier
\section{No-timeslice approach}
\label{sec-no-timeslice}

Finally, a couple of words may be said on the case when there is no need to
study the intermediate stages of the neutron spectrum (its time slices), but
there is still interest in the form of a stationary spectrum. In this case the
algorithm described in Section~\ref{sec-timeslices-Geant4} may be further
simplified as follows.

Since each of the \texttt{PreStepPoint} and \texttt{PostStepPoint} provide the
information on particle's \texttt{GlobalTime} $t$ and kinetic energy $E$, it is
easy to calculate the particle's contribution to the corresponding energy bin
in the total stationary spectrum:

\begin{equation}
p(E) = \frac{t_{POST} - t_{PRE}}{T_{max}}
\end{equation}

This method does not require the introduction of the concept of \textit{time
slices} described above, and is therefore free of its accompanying
complications. While it is still usable for calculating the final stationary
spectrum, there are many cases when it will not suffice, because the stages of
spectrum evolution may be of considerable interest.

\begin{figure}[h!]
\begin{center}
\includegraphics[width=14cm]{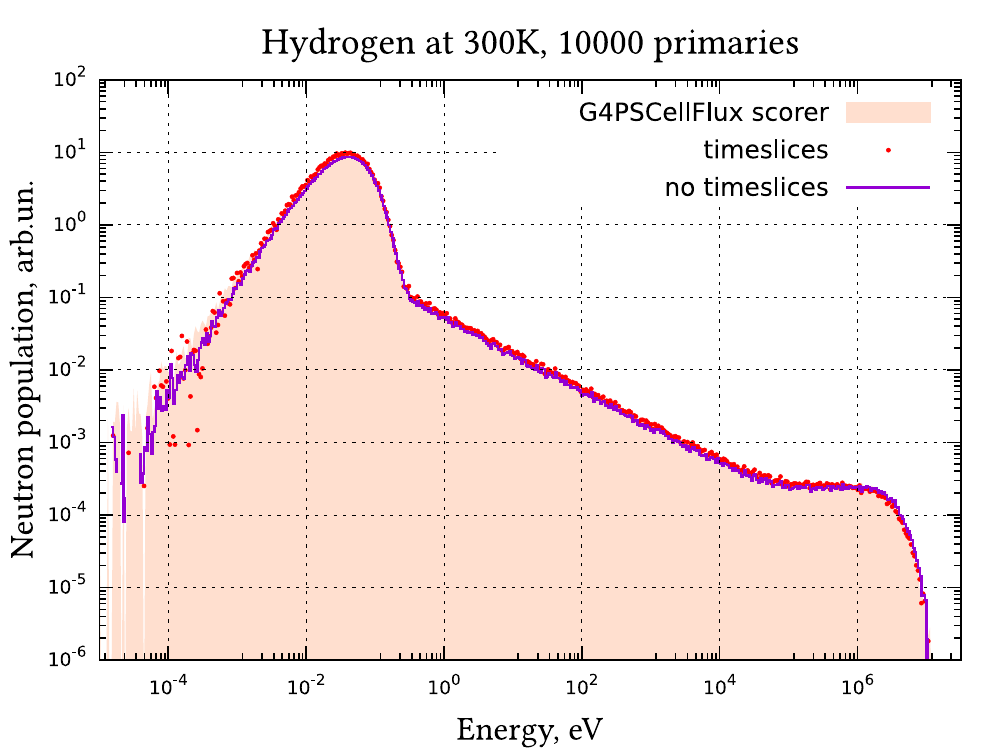}
\end{center}
\caption{Stationary neutron spectra calculated through the time slices (red
         dots) and widthout the time slices (violet line). As can be seen, with
         sufficiently high number of primary particles, these curves coincide
         very closely. The random fluctuations in the low energy range here are more prominent than in Figs.~\ref{fig-total-spectrum-lintime} and~\ref{fig-total-spectrum-Hydrogen-logtime} because of a much smaller number of primaries simulated, while the bin width in the lower energy range is orders of magnitude smaller than that in the higher energy range (logarithmic bin width described in Section~\ref{sec-timeslices-Geant4}, which is why the adjacent bins are filled very unevenly).}
\label{fig-timeslices-notimeslices}
\end{figure}

An example of stationary neturon spectrum calculated this way is shown in 
Fig.~\ref{fig-timeslices-notimeslices} compared to the one obtained with time
slices approach (Fig.~\ref{fig-total-spectrum-Hydrogen-logtime}).

As can be seen in this figure, the steady state spectra calculated using both methods (with time slices and without them) are practically the same even for a rather small number of primary events generated (orders of magnitude smaller than those used to produce Figs.~\ref{fig-total-spectrum-lintime} and~\ref{fig-total-spectrum-Hydrogen-logtime}). This simple method without time slices, described in this section, is also roughly equivalent to the use of the built-in \texttt{CellFlux} scorer in Geant4. Indeed, to retrieve a neutron spectrum, one can create a number of identical \texttt{CellFlux} scorers and attach corresponding particle and kinetic energy filters to each of them (a scorer per each energy bin)~\mbox{\cite{MakotoAsaiScoring2019}}. Then, after the simulation, one can collect the values accumulated by all scorers to obtain a complete spectrum. Aside from the minor inconvenience of creating and operating the numerous scorers, the \texttt{CellFlux} scorer does not yet support attaching to real-world logical volumes (as of Geant4 version 11.1), so in order to use it, one has also to create a dedicated scoring mesh to fit the geometry of their detector. Recent versions of Geant4 (starting with 10.7) feature some classes for easier accumulation of histograms. In particular, with the new \texttt{G4TScoreHistFiller} class, a primitive scorer can directly fill a 1-D histogram. This spares a user the necessity to deal with multiple scorers and collect all the values at the end of simulation, but also comes with its own limitations. For example, one can only use this histogram filler with the ``real-world'' or ``probe'' scorers. Thus, the only possible way of utilizing the \texttt{CellFlux} scorer together with \texttt{ScoreHistFiller} at the moment is to attach them to a ``probe'' (which is essentially a cube). This was exactly the way we obtained the data shown with light orange shaded area in Fig.~\mbox{\ref{fig-timeslices-notimeslices}} for comparison. It should be noted that we had to limit the maximum length of a single step to 1~mm for neutrons in order to achieve the accurate results for the hydrogen at normal conditions.

The comparison shows that the form of the steady-state spectrum is exactly the same, but the downside of the approach with built-in \texttt{CellFlux} scorer is the impossibility to attach it directly to a logical volume of arbitrary form, which would obviously complicate things in more realistic use cases.
Additionally, both the ``timeslices'' and ``no timeslices'' curves in Fig.~\mbox{\ref{fig-timeslices-notimeslices}} 
were calculated much faster than the one obtained with the built-in \texttt{CellFlux} scorer and a histogram filler on the same test machine.

And again, all the methods described in this section give absolutely no information about the intermediate shapes of the neutron spectrum, as opposed to our first, ``time-slices approach'' described in Section~\ref{sec-timeslices-Geant4} above. Meanwhile, these intermediate ``time slices'' or ``snapshots'' of the neutron spectrum carry a lot of useful information for future studies (see e.g.~\cite{Smolyar2021JPS}).

\section{Optimization considerations}
\label{sec-optimization}

\subsection{Parallelization}

\subsubsection{Shared memory parallelization}

In order to utilize the shared memory parallelism in Geant4, one should first
instantiate the \texttt{G4MTRunManager} instead of \texttt{G4RunManager}. This
run manager creates several instances of the \texttt{G4Run} class and divides
the task between them. The \texttt{SetNumberOfThreads()} method may be used to
set the desired number of threads. Each instance of G4Run will collect the data
independently, so it is convenient to make the \texttt{pTimeSpectrum} array 
a member of its subclass. At the end of the job the data collected by all
instances of \texttt{G4Run} must be merged together. For this purpose the 
\texttt{G4MTRunManager} calls the master run's \texttt{Merge()} method with
each of them. The spectrum arrays may be simply added like this:
%
%
\begin{sourcecode}{}
void Run::Merge(const G4Run* otherRun ) 
{
  const Run* other = static_cast<const Run*>(otherRun);

  for (int i = 0; i < numEnergyBins; i++)
    for (int j = 0; j < numTimeBins; j++)
      pTimeSpectrum[i][j] += other->pTimeSpectrum[i][j];

  G4Run::Merge(other);
}
\end{sourcecode}

After that the \texttt{RunAction::EndOfRunAction()} method can output the
resulting arrays for the master run, which may be checked with the 
\texttt{isMaster} variable.

\subsubsection{Distributed memory parallelization}

The Geant4 simulation can also be parallelized over the distributed memory
system, e.g. several computers with network connection. A \texttt{G4MPImanager}
from Geant4's extended examples can be used for 
this\footnote{geant4.10.07/examples/extended/parallel/MPI/}. First, it is
necessary to open the MPI session just before instantiating the
\texttt{G4RunManager} or \texttt{G4MTRunManager}:
\begin{sourcecode}{}
G4MPImanager *g4MPI = new G4MPImanager(argc,argv);
G4MPIsession *session = g4MPI->GetMPIsession();

//G4RunManager *runManager = new G4RunManager;
G4MTRunManager *runManager = new G4MTRunManager;
\end{sourcecode}{}

In this case the MPI workload manager or the directly invoked \texttt{mpirun}
utility will launch a separate Geant4 application on each machine with its own
run manager.

At the end of the job the collected data must be sent over the network to the
MPI process with rank 0. For this purpose, the mentioned Geant4 example
provides the \texttt{G4VUserMPIrunMerger} class, which has the \texttt{Pack()}
and \texttt{UnPack()} methods to send the collected data over network as a
single 1D array (hence the names). The inherited \texttt{RunMerger} class must
be instantiated and activated in the \texttt{RunAction::EndOfRunAction()}
method: 
\begin{sourcecode}{}
RunMerger rm(static_cast<const Run*>(currentRun));
rm.Merge();
\end{sourcecode}

Although it is possible to run multiple MPI sessions on a single multi-core
machine and speed-up the lengthy calculations that way, there is actually a
more optimal way. As shown in the example above, the local run manager may
also be multithreaded. In this case, the \texttt{RunMerger::Merge()} should be
called by the local master runs only, after all the local worker runs are
merged as described in the previous section. This way it is possible to benefit
from the distributed memory MPI parallelization while saving the MPI data
transfer overhead for the local shared memory threads.

\subsection{Pre-caclculated cross-sections}

Another way to speed-up the Geant4 simulation involving neutrons is to supply
the pre-calculated nuclear reaction cross-sections for the desired temperature.
The data on the incident neutron cross-sections may be found e.g. in 
{ENDF/B-VIII.0} library~\cite{ENDFBVIII0,ENDFBVIII0paper}. However, the 
cross-sections must be recalculated for the required temperature. This can
be done using the PREPRO codes~\cite{PREPRO2019} in the following sequence:
\begin{tcolorbox}[]
Endf2c $\rightarrow$ Linear $\rightarrow$ Recent $\rightarrow$ Sigma1
$\rightarrow$ Fixup $\rightarrow$ Dictin
\end{tcolorbox}
\noindent which produces the recalculated cross-sections for all available
neutron reactions in the ENDF6 format, same as input. The cross-sections for
certain reactions may then be conveniently extracted separately using the
\texttt{pyENDF6} module\footnote{\url{https://github.com/DavidWalz/pyENDF6}}.
Thus obtained data must be loaded and stored in some \texttt{ExternalXSDataset}
class inheriting from the Geant4's\linebreak \texttt{G4VCrossSectionDataSet} class and
implementing the \texttt{IsIsoApplicable()} and\linebreak \texttt{GetIsoCrossSection()}
methods. These datasets must be instantiated in the\linebreak 
\texttt{PhysicsList::ConstructProcess()} and attached to each process of
interest:
\begin{sourcecode}{}
void PhysicsList::ConstructProcess() {
  AddTransportation();
  G4PhysConstVector *physicsVector = 
    (G4VMPLsubInstanceManager.offset[g4vmplInstanceID]).physicsVector;
  G4PhysConstVector::iterator itr;
  for (itr = physicsVector->begin(); itr != physicsVector->end(); ++itr) 
    (*itr)->ConstructProcess();

  G4ProcessManager *pmanager = G4Neutron::Neutron()->GetProcessManager();
  G4ProcessVector *processVector = pmanager->GetProcessList();
  G4String processName;
  for (int i=0; i < processVector->entries(); i++)
  {
    processName = processVector[i]->GetProcessName();
    if (processName == "hadElastic")
    {
      G4HadronElasticProcess *process = 
                             (G4HadronElasticProcess *) processVector[i];
      ExternalXSDataset *H1Elastic = new ExternalXSDataset(ELASTIC); 
      process->GetCrossSectionDataStore()->AddDataSet(H1Elastic);
    }
    else if (processName == "nCapture")
    {
      G4HadronCaptureProcess *process = 
                             (G4HadronCaptureProcess *) processVector[i];
      ExternalXSDataset *H1Capture = new ExternalXSDataset(CAPTURE);
      process->GetCrossSectionDataStore()->AddDataSet(H1Capture);
    }
  }
}
\end{sourcecode}

It is then the user's responsibility to supply the correct cross-sections for
the specified temperature.

\section{Conclusions}

We considered the case of neutron moderation under the action of a fixed
neutron source. The knowledge of the non-equilibrium stationary neutron
spectrum is essential to a number of neutron-related physical problems.

We developed an algorithm of the stationary spectrum calculation based on the
concept of spectrum ``time slices'', which provides the view of the spectrum
evolution stages
for a single ``generation'' of neutrons, as well as for the
whole set of neutrons including all new generations coming from the constantly
acting neutron source. We also showed a way to implement this algorithm in the
custom software built on top of the Geant4 Monte Carlo simulation toolkit.

The developed software algorithm was tested against the known neutron
moderation spectrum in hydrogen and demonstrated a good agreement. We also
illustrated the advantages of the use of logarithmic time scale in such
problems and suggested a simple and quick way of its implementation.

Since the described algorithm and its implementation are rather universal, this
software may be used for various particular tasks with only minor
modifications (as small as change the system geometry and the neutron source
spectrum). Despite the fact that we tested the software for the infinite
homogeneous medium only, it may as well be applied for any kind of
heterogeneous structures.

\section*{Acknowledgements}

The authors wish to thank the anonymous peer reviewers for a fruitful discussion and very useful hints, which, among other things, helped the authors to extend their knowledge and the toolset with some new techniques.

\bibliographystyle{elsarticle-num} 
\bibliography{Smolyar-Geant4-moderation}

\begin{thebibliography}{10}
\expandafter\ifx\csname url\endcsname\relax
  \def\url#1{\texttt{#1}}\fi
\expandafter\ifx\csname urlprefix\endcsname\relax\def\urlprefix{URL }\fi
\expandafter\ifx\csname href\endcsname\relax
  \def\href#1#2{#2} \def\path#1{#1}\fi

\bibitem{Ferziger1966}
J.~F. Ferziger, P.~F. Zweifel, The Theory of Neutron Slowing Down in Nuclear
  Reactors, M.I.T. Press, 1966.

\bibitem{Williams1966}
M.~M.~R. Williams, The Slowing Down and Thermalization of Neutrons,
  North-Holland, Amsterdam, 1966.

\bibitem{Stacey2018}
W.~M. Stacey, Nuclear Reactor Physics, 3rd Edition, Wiley, 2018.
\newblock \href {https://doi.org/10.1002/9783527812318}
  {\path{doi:10.1002/9783527812318}}.

\bibitem{Rusov2015Peculiarities}
V.~D. Rusov, V.~A. Tarasov, I.~V. Sharph, V.~N. Vashchenko, E.~P. Linnik, T.~N.
  Zelentsova, M.~E. Beglaryan, S.~A. Chernegenko, S.~I. Kosenko, V.~P. Smolyar,
  On some fundamental peculiarities of the traveling wave reactor, Science and
  Technology of Nuclear Installations (2015) 703069\href
  {https://doi.org/10.1155/2015/703069} {\path{doi:10.1155/2015/703069}}.

\bibitem{Feoktistov1989}
L.~Feoktistov, Neutron-fission wave, Dokl. Akad. Nauk SSSR~(309) (1989) 4--7,
  (in Russian).

\bibitem{Teller1996}
E.~Teller, M.~Ishikawa, L.~Wood, R.~Hyde, J.~Nuckolls, Completely automated
  nuclear reactors for long-term operation ii: Toward a concept-level
  point-desig n of a high-temperature, gas-cooled central power station system,
  part ii, in: Proceedings of the International Conference on Emerging Nuclear
  Energy Systems, ICENES'96, Obninsk, Russian Federation, Obninsk, Russian
  Federation, Obninsk, Russian Federation, 1996, pp. 123--127, also available
  from Lawrence Livermore National Laboratory, California, publication
  UCRL-JC-122708-R T2.

\bibitem{Rusov2011Energies}
V.~D. Rusov, E.~P. Linnik, V.~A. Tarasov, T.~N. Zelentsova, I.~V. Sharph, V.~N.
  Vaschenko, S.~I. Kosenko, M.~E. Beglaryan, S.~A. Chernezhenko, P.~A.
  M~olchinikolov, S.~I. Saulenko, O.~A. Byegunova, Traveling wave reactor and
  condition of existence of nuclear burning soliton-like wave in
  neutron-multiply ing media, Energies 4~(9) (2011) 1337--1361.
\newblock \href {https://doi.org/10.3390/en4091337}
  {\path{doi:10.3390/en4091337}}.

\bibitem{Tarasov2015Ultraslow}
V.~Rusov, V.~Tarasov, M.~Eingorn, S.~Chernezhenko, A.~Kakaev, V.~Vashchenko,
  M.~Beglaryan, Ultraslow wave nuclear burning of uranium–plutonium fissile
  medium on epithermal neutrons, Progress in Nuclear Energy 83 (2015) 105 --
  122.
\newblock \href {https://doi.org/10.1016/j.pnucene.2015.03.007}
  {\path{doi:10.1016/j.pnucene.2015.03.007}}.

\bibitem{Rusov2011Sharpening}
V.~Rusov, V.~Tarasov, S.~Chernezhenko, The modes with the sharpening in the
  uranium-plutonium fission environment of the technical nuclear reactors and
  georeactor, Problems of atomic science and technology 72 (2011) 123--131, (in
  Russian).

\bibitem{Rusov2013Fukushima}
V.~Rusov, V.~Tarasov, V.~Vaschenko, E.~Linnik, T.~Zelentsova, M.~Beglaryan,
  S.~Chernegenko, S.~Kosenko, P.~Molchinikolov, V.~Smolyar, E.~Grechan,
  Fukushima plutonium effect and blow-up regimes in neutron-multiplying media,
  World Journal of Nuclear Science and Technology 3~(2A) (2013) 9--18.
\newblock \href {https://doi.org/10.4236/wjnst.2013.32A002}
  {\path{doi:10.4236/wjnst.2013.32A002}}.

\bibitem{MacFarlane2010}
R.~E. MacFarlane, 3. Neutron Slowing Down and Thermalization, Springer US,
  Boston, MA, 2010, pp. 189--277.
\newblock \href {https://doi.org/10.1007/978-0-387-98149-9_3}
  {\path{doi:10.1007/978-0-387-98149-9_3}}.

\bibitem{Kok2016}
K.~D. Kok (Ed.), Nuclear Engineering Handbook, 2nd Edition, CRC Press, 2016.
\newblock \href {https://doi.org/10.1201/9781315373829}
  {\path{doi:10.1201/9781315373829}}.

\bibitem{Deiev2013}
O.~Deiev, {GEANT4} simulation of neutron transport and scattering in media,
  Problems of atomic science and technology 85 (2013) 236--241.

\bibitem{Lisovska2019}
V.~Lisovska, T.~Malykhina, V.~Shpagina, R.~Timchenko, {GEANT4} modeling of
  energy spectrum of fast neutrons source for the development of research
  technique of heavy scintillators, East European Journal of Physics~(2) (2019)
  58--63.
\newblock \href {https://doi.org/10.26565/2312-4334-2019-2-09}
  {\path{doi:10.26565/2312-4334-2019-2-09}}.

\bibitem{Shin2014}
J.~W. Shin, S.-W. Hong, S.-I. Bak, D.~Y. Kim, C.~Y. Kim, {GEANT4} and {PHITS}
  simulations of the shielding of neutrons from the {252Cf} source, Journal of
  the Korean Physical Society 65~(5) (2014) 591–598.
\newblock \href {https://doi.org/10.3938/jkps.65.591}
  {\path{doi:10.3938/jkps.65.591}}.

\bibitem{Robinson2014}
A.~E. Robinson, New libraries for simulating neutron scattering in dark matter
  detector calibrations, Physical Review C 89~(3) (Mar 2014).
\newblock \href {https://doi.org/10.1103/physrevc.89.032801}
  {\path{doi:10.1103/physrevc.89.032801}}.

\bibitem{Lemrani2006}
R.~Lemrani, M.~Robinson, V.~Kudryavtsev, M.~De~Jesus, G.~Gerbier, N.~Spooner,
  Low-energy neutron propagation in {MCNPX} and {GEANT4}, Nuclear Instruments
  and Methods in Physics Research Section A: Accelerators, Spectrometers,
  Detectors and Associated Equipment 560~(2) (2006) 454–459.
\newblock \href {https://doi.org/10.1016/j.nima.2005.12.238}
  {\path{doi:10.1016/j.nima.2005.12.238}}.

\bibitem{GEANT4DevBook2022}
{Geant4 Collaboration}, {GEANT4}: Book for application developers. release
  11.1, rev7.0, https://geant4.web.cern.ch/docs/ (December 2022).

\bibitem{Geant4Toolkit2003}
S.~Agostinelli, J.~Allison, K.~Amako, J.~Apostolakis, H.~Araujo, P.~Arce,
  M.~Asai, D.~Axen, S.~Banerjee, G.~Barrand, F.~Behner, L.~Bellagamba,
  J.~Boudreau, L.~Broglia, A.~Brunengo, H.~Burkhardt, S.~Chauvie, J.~Chuma,
  R.~Chytracek, G.~Cooperman, G.~Cosmo, P.~Degtyarenko, A.~Dell'Acqua,
  G.~Depaola, D.~Dietrich, R.~Enami, A.~Feliciello, C.~Ferguson, H.~Fesefeldt,
  G.~Folger, F.~Foppiano, A.~Forti, S.~Garelli, S.~Giani, R.~Giannitrapani,
  D.~Gibin, J.~G. Cadenas, I.~Gonz\'{a}lez, G.~G. Abril, G.~Greeniaus,
  W.~Greiner, V.~Grichine, A.~Grossheim, S.~Guatelli, P.~Gumplinger,
  R.~Hamatsu, K.~Hashimoto, H.~Hasui, A.~Heikkinen, A.~Howard, V.~Ivanchenko,
  A.~Johnson, F.~Jones, J.~Kallenbach, N.~Kanaya, M.~Kawabata, Y.~Kawabata,
  M.~Kawaguti, S.~Kelner, P.~Kent, A.~Kimura, T.~Kodama, R.~Kokoulin,
  M.~Kossov, H.~Kurashige, E.~Lamanna, T.~Lamp\'{e}n, V.~Lara, V.~Lefebure,
  F.~Lei, M.~Liendl, W.~Lockman, F.~Longo, S.~Magni, M.~Maire, E.~Medernach,
  K.~Minamimoto, P.~M. de~Freitas, Y.~Morita, K.~Murakami, M.~Nagamatu,
  R.~Nartallo, P.~Nieminen, T.~Nishimura, K.~Ohtsubo, M.~Okamura, S.~O'Neale,
  Y.~Oohata, K.~Paech, J.~Perl, A.~Pfeiffer, M.~Pia, F.~Ranjard, A.~Rybin,
  S.~Sadilov, E.~D. Salvo, G.~Santin, T.~Sasaki, N.~Savvas, Y.~Sawada,
  S.~Scherer, S.~Sei, V.~Sirotenko, D.~Smith, N.~Starkov, H.~Stoecker,
  J.~Sulkimo, M.~Takahata, S.~Tanaka, E.~Tcherniaev, E.~S. Tehrani,
  M.~Tropeano, P.~Truscott, H.~Uno, L.~Urban, P.~Urban, M.~Verderi, A.~Walkden,
  W.~Wander, H.~Weber, J.~Wellisch, T.~Wenaus, D.~Williams, D.~Wright,
  T.~Yamada, H.~Yoshida, D.~Zschiesche, Geant4 -- a simulation toolkit, Nuclear
  Instruments and Methods in Physics Research Section A: Accelerators,
  Spectrometers, Detectors and Associated Equipment 506~(3) (2003) 250--303.
\newblock \href {https://doi.org/10.1016/S0168-9002(03)01368-8}
  {\path{doi:10.1016/S0168-9002(03)01368-8}}.

\bibitem{Geant4applications2006}
J.~{Allison}, K.~{Amako}, J.~{Apostolakis}, H.~{Araujo}, P.~{Arce Dubois},
  M.~{Asai}, G.~{Barrand}, R.~{Capra}, S.~{Chauvie}, R.~{Chytracek}, G.~A.~P.
  {Cirrone}, G.~{Cooperman}, G.~{Cosmo}, G.~{Cuttone}, G.~G. {Daquino},
  M.~{Donszelmann}, M.~{Dressel}, G.~{Folger}, F.~{Foppiano}, J.~{Generowicz},
  V.~{Grichine}, S.~{Guatelli}, P.~{Gumplinger}, A.~{Heikkinen},
  I.~{Hrivnacova}, A.~{Howard}, S.~{Incerti}, V.~{Ivanchenko}, T.~{Johnson},
  F.~{Jones}, T.~{Koi}, R.~{Kokoulin}, M.~{Kossov}, H.~{Kurashige}, V.~{Lara},
  S.~{Larsson}, F.~{Lei}, O.~{Link}, F.~{Longo}, M.~{Maire}, A.~{Mantero},
  B.~{Mascialino}, I.~{McLaren}, P.~{Mendez Lorenzo}, K.~{Minamimoto},
  K.~{Murakami}, P.~{Nieminen}, L.~{Pandola}, S.~{Parlati}, L.~{Peralta},
  J.~{Perl}, A.~{Pfeiffer}, M.~G. {Pia}, A.~{Ribon}, P.~{Rodrigues},
  G.~{Russo}, S.~{Sadilov}, G.~{Santin}, T.~{Sasaki}, D.~{Smith}, N.~{Starkov},
  S.~{Tanaka}, E.~{Tcherniaev}, B.~{Tome}, A.~{Trindade}, P.~{Truscott},
  L.~{Urban}, M.~{Verderi}, A.~{Walkden}, J.~P. {Wellisch}, D.~C. {Williams},
  D.~{Wright}, H.~{Yoshida}, Geant4 developments and applications, IEEE
  Transactions on Nuclear Science 53~(1) (2006) 270--278.
\newblock \href {https://doi.org/10.1109/TNS.2006.869826}
  {\path{doi:10.1109/TNS.2006.869826}}.

\bibitem{Allison2016}
J.~Allison, K.~Amako, J.~Apostolakis, P.~Arce, M.~Asai, T.~Aso, E.~Bagli,
  A.~Bagulya, S.~Banerjee, G.~Barrand, B.~Beck, A.~Bogdanov, D.~Brandt,
  J.~Brown, H.~Burkhardt, P.~Canal, D.~Cano-Ott, S.~Chauvie, K.~Cho,
  G.~Cirrone, G.~Cooperman, M.~Cortés-Giraldo, G.~Cosmo, G.~Cuttone,
  G.~Depaola, L.~Desorgher, X.~Dong, A.~Dotti, V.~Elvira, G.~Folger,
  Z.~Francis, A.~Galoyan, L.~Garnier, M.~Gayer, K.~Genser, V.~Grichine,
  S.~Guatelli, P.~Guèye, P.~Gumplinger, A.~Howard, I.~Hřivnáčová,
  S.~Hwang, S.~Incerti, A.~Ivanchenko, V.~Ivanchenko, F.~Jones, S.~Jun,
  P.~Kaitaniemi, N.~Karakatsanis, M.~Karamitros, M.~Kelsey, A.~Kimura, T.~Koi,
  H.~Kurashige, A.~Lechner, S.~Lee, F.~Longo, M.~Maire, D.~Mancusi, A.~Mantero,
  E.~Mendoza, B.~Morgan, K.~Murakami, T.~Nikitina, L.~Pandola, P.~Paprocki,
  J.~Perl, I.~Petrović, M.~Pia, W.~Pokorski, J.~Quesada, M.~Raine, M.~Reis,
  A.~Ribon, A.~R. Fira], F.~Romano, G.~Russo, G.~Santin, T.~Sasaki, D.~Sawkey,
  J.~Shin, I.~Strakovsky, A.~Taborda, S.~Tanaka, B.~Tomé, T.~Toshito, H.~Tran,
  P.~Truscott, L.~Urban, V.~Uzhinsky, J.~Verbeke, M.~Verderi, B.~Wendt,
  H.~Wenzel, D.~Wright, D.~Wright, T.~Yamashita, J.~Yarba, H.~Yoshida, Recent
  developments in {Geant4}, Nuclear Instruments and Methods in Physics Research
  Section A: Accelerators, Spectrometers, Detectors and Associated Equipment
  835 (2016) 186 -- 225.
\newblock \href {https://doi.org/https://doi.org/10.1016/j.nima.2016.06.125}
  {\path{doi:https://doi.org/10.1016/j.nima.2016.06.125}}.

\bibitem{Tarasov2017Moderation}
V.~D. Rusov, V.~A. Tarasov, S.~A. Chernezhenko, A.~A. Kakaev, V.~P. Smolyar,
  Neutron moderation theory with thermal motion of the moderator nuclei, The
  European Physical Journal A 53~(9) (2017) 179.
\newblock \href {https://doi.org/10.1140/epja/i2017-12363-9}
  {\path{doi:10.1140/epja/i2017-12363-9}}.

\bibitem{GEANT4PhysRef2017}
{Geant4 Collaboration}, {GEANT4}: Physics reference manual. release 10.4,
  rev1.0, http://geant4.web.cern.ch/support (December 2017).

\bibitem{Apostolakis2009}
J.~Apostolakis, M.~Asai, A.~Bogdanov, H.~Burkhardt, G.~Cosmo, S.~Elles,
  G.~Folger, V.~Grichine, P.~Gumplinger, A.~Heikkinen, I.~Hrivnacova,
  V.~Ivanchenko, J.~Jacquemier, T.~Koi, R.~Kokoulin, M.~Kossov, H.~Kurashige,
  I.~McLaren, O.~Link, M.~Maire, W.~Pokorski, T.~Sasaki, N.~Starkov, L.~Urban,
  D.~Wright, Geometry and physics of the {Geant4} toolkit for high and medium
  energy applications, Radiation Physics and Chemistry 78~(10) (2009) 859 --
  873, workshop on Use of Monte Carlo Techniques for Design and Analysis of
  Radiation Detectors.
\newblock \href
  {https://doi.org/https://doi.org/10.1016/j.radphyschem.2009.04.026}
  {\path{doi:https://doi.org/10.1016/j.radphyschem.2009.04.026}}.

\bibitem{Wright2015}
D.~Wright, M.~Kelsey, The {Geant4} bertini cascade, Nuclear Instruments and
  Methods in Physics Research Section A: Accelerators, Spectrometers, Detectors
  and Associated Equipment 804 (2015) 175 -- 188.
\newblock \href {https://doi.org/https://doi.org/10.1016/j.nima.2015.09.058}
  {\path{doi:https://doi.org/10.1016/j.nima.2015.09.058}}.

\bibitem{G4NDL}
{G4NDL} -- neutron data files with thermal cross sections,
  http://cern.ch/geant4-data/datasets/G4NDL.4.5.tar.gz.

\bibitem{Oka2014}
Y.~Oka (Ed.), Nuclear Reactor Design, Springer, 2014.
\newblock \href {https://doi.org/10.1007/978-4-431-54898-0}
  {\path{doi:10.1007/978-4-431-54898-0}}.

\bibitem{Smolyar2021JPS}
V.~P. Smolyar, A.~O. Mileva, V.~O. Tarasov, H.~H. Neboha, V.~D. Rusov,
  Simulation of the neutron spectra evolution with {GEANT4} {Monte} {Carlo}
  code, Journal of Physical Studies 25~(2) (2021) 2201, (in Ukrainian).
\newblock \href {https://doi.org/10.30970/jps.25.2201}
  {\path{doi:10.30970/jps.25.2201}}.

\bibitem{MakotoAsaiScoring2019}
M.~Asai, Scoring {I} -- {Geant4 Tutorial Course}, version 10.5, {First}
  {Geant4} {Tutorial} {Workshop} at {Instituto} de {F\'{i}sica} da
  {Universidade} de {S\~{a}o} {Paulo},
  \url{https://indico.cern.ch/event/776050/contributions/3237924/} (2019).

\bibitem{ENDFBVIII0}
{ENDF/B-VIII.0} incident-neutron data,
  \url{https://www.nndc.bnl.gov/endf/b8.0/download.html}.

\bibitem{ENDFBVIII0paper}
D.~Brown, M.~Chadwick, R.~Capote, A.~Kahler, A.~Trkov, M.~Herman, A.~Sonzogni,
  Y.~Danon, A.~Carlson, M.~Dunn, D.~Smith, G.~Hale, G.~Arbanas, R.~Arcilla,
  C.~Bates, B.~Beck, B.~Becker, F.~Brown, R.~Casperson, J.~Conlin, D.~Cullen,
  M.-A. Descalle, R.~Firestone, T.~Gaines, K.~Guber, A.~Hawari, J.~Holmes,
  T.~Johnson, T.~Kawano, B.~Kiedrowski, A.~Koning, S.~Kopecky, L.~Leal,
  J.~Lestone, C.~Lubitz, J.~{M\'{a}rquez Dami\'{a}n}, C.~Mattoon, E.~McCutchan,
  S.~Mughabghab, P.~Navratil, D.~Neudecker, G.~Nobre, G.~Noguere, M.~Paris,
  M.~Pigni, A.~Plompen, B.~Pritychenko, V.~Pronyaev, D.~Roubtsov, D.~Rochman,
  P.~Romano, P.~Schillebeeckx, S.~Simakov, M.~Sin, I.~Sirakov, B.~Sleaford,
  V.~Sobes, E.~Soukhovitskii, I.~Stetcu, P.~Talou, I.~Thompson, S.~{van der
  Marck}, L.~Welser-Sherrill, D.~Wiarda, M.~White, J.~Wormald, R.~Wright,
  M.~Zerkle, G.~\v{Z}erovnik, Y.~Zhu, {ENDF/B-VIII.0}: The 8th major release of
  the nuclear reaction data library with cielo-project cross sections, new
  standards and thermal scattering data, Nuclear Data Sheets 148 (2018) 1--142,
  special Issue on Nuclear Reaction Data.
\newblock \href {https://doi.org/https://doi.org/10.1016/j.nds.2018.02.001}
  {\path{doi:https://doi.org/10.1016/j.nds.2018.02.001}}.

\bibitem{PREPRO2019}
D.~Cullen, \href{https://www-nds.iaea.org/public/endf/prepro/}{{PREPRO} 2019:
  2019 {ENDF/B} pre-processing codes}, Tech. Rep. IAEA-NDS-39, Rev. 19, IAEA
  (2019).
\newline\urlprefix\url{https://www-nds.iaea.org/public/endf/prepro/}

\end{thebibliography}

\end{document}